\documentclass[twocolumn,prl,showpacs,preprintnumbers,amsmath,amssymb]{revtex4}

\usepackage{graphicx}
\usepackage{dcolumn}
\usepackage{bm}

\begin{document}

\preprint{APS/123-QED}

\title{Binary Quantum Turbulence Arising from Countersuperflow Instability\\
 in Two-Component Bose-Einstein Condensates}

\author{Hiromitsu Takeuchi, Shungo Ishino, and Makoto Tsubota}
\affiliation{%
Department of Physics, Osaka City University, Sumiyoshi-ku, Osaka 558-8585, Japan 
}%


\date{\today}

\begin{abstract} 
We theoretically study the development of quantum turbulence from two counter-propagating superfluids
 of miscible Bose--Einstein condensates by numerically solving the coupled Gross--Pitaevskii equations.
 When the relative velocity exceeds a critical value, the countersuperflow becomes unstable and quantized
 vortices are nucleated, which leads to isotropic quantum turbulence consisting of two superflows.
 It is shown that the binary turbulence can be realized experimentally in a trapped system.
\end{abstract}


\maketitle
Hydrodynamic instability develops into turbulence or turbulent flow due to the complicated dynamics of eddies of various length scales \cite{Kund}.
 This phenomenon appears in diverse fields of physics,
 such as magneto-hydrodynamic fluids in plasma physics \cite{Hasegawa},
 relativistic fluids in astrophysics \cite{Perucho},
 and quantum fluids in low temperature physics \cite{Takeuchi, Sasaki}.
Turbulence has been studied thoroughly in classical fluid dynamics as the most important unsolved problem of classical physics.
 Recently, quantum turbulence (QT), turbulence in superfluids, has attracted interest in low temperature physics \cite{PLTP}.
 QT consists of topological defects, vortices with definite quantized circulation, and can provide a simple prototype for understanding classical turbulence.
 QT has been historically studied in superfluid $^4$He and $^3$He
 and has been numerically studied in atomic Bose--Einstein condensates (BECs) \cite{Kobayashi07, White} and has recently been realized in experiments \cite{Henn09}.

Early studies on QT were devoted primarily to thermal counterflow in superfluid $^4$He,
 where the hydrodynamics is usually described using a two-fluid model in which the system consists of an inviscid superfluid component and a viscous normal fluid component \cite{Khalatnikov1}.
 Thermal counterflow is driven by an injected heat current,
 where the two components flow in opposite directions.
 For a sufficiently large relative velocity between the two components,
 the laminar counterflow state becomes unstable, developing into a superfluid turbulent state.
 In the transition,
 remnant vortices, quantized vortices attached to the container wall, are stretched by the mutual friction between the superfluid and normal fluid components,
 growing into a tangle through reconnections with other vortices \cite{DonnellyBook, Adachi}. 

 We consider a similar case in two-component BECs,
 QT arising from the instability of the counterflow of two superfluids, namely, countersuperflow instability.
 Counter-superflow instability has been theoretically studied in several multicomponent miscible superfluid systems, such as 
 helium superfluids \cite{Khalatnikov2,Mineev,Nepomnyashchii, Yukalov},
 mixture BECs of cold atoms \cite{Law},
 and nucleon superflows in rotating neutron stars \cite{Andersson}.
  Atomic BEC systems have great advantages over other superfluid systems,
 because modern experimental techniques enable us to easily control superflows and directly visualize the dynamics of topological defects therein.
 Recently, Hamner {\it et al.} experimentally realized countersuperflow instability for the first time,
 and observed that shocks and dark-bright solitons nucleate via the instability in quasi-one-dimensional two-component BECs \cite{Hamner10}.
 We expect that countersuperflow instability exhibits novel hydrodynamic instability as was reported in \cite{Takeuchi, Sasaki}, and causes quantum turbulence in two-component BECs as in the case of the thermal counterflow turbulence. 
 In the context of the research of study QT, it is remarkable that the superfluid velocity field in two-component BECs is continuous in contrast to that in single-component superfluids \cite{Kasamatsu05}.
 The continuous velocity field enable us to calculate the important statistical quantities in classical turbulence such as enstrophy and helicity,
 which are not well defined in single-component superfluid systems.  
 Hence we expect that countersuperflow instability in two-component BECs will open new directions for research into unique turbulent states in multicomponent superfluids.

\begin{figure*}
\centering
 \includegraphics[width=.7 \linewidth]{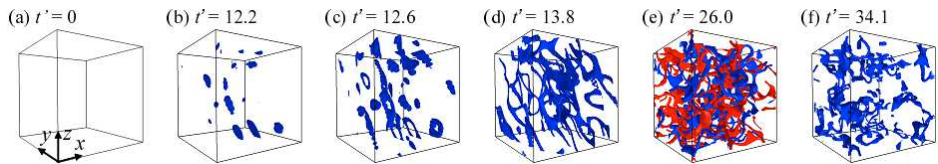}
 \caption{ (Color online)
Nonlinear dynamics of countersuperflow instability.
 The figures show the isosurfaces of the density $n_1=0.05n_0$ of the $1$st component.
 The instability causes (b) disk-shaped density pulses which (c) transform into vortex rings.
 The vortex rings are enlarged by the momentum exchange between the two condensates,
 and then (d) the vortex interactions and reconnections deform the rings,
 leading to (e) binary quantum turbulence.
 (f) Then the vortex tangle decays.
 The isosurfaces with $n_2=0.05n_0$ of the 2nd component is also plotted in (e) in a different color.
 We solved the GP equations with the relative velocity $U_{12}=4.71 \times \sqrt{gn_0/m}$ on $64^3$ grid.
 The box size is $(11\times\xi_v)^3$ with the characteristic vortex core size $\xi_v=\hbar/\sqrt{mn_0(g+g_{12})}$.
 Time $t$ is normalized as $t'=gn_0t/\hbar$.
}
\label{fig:dynamics}
\end{figure*}

In this work we investigate the development of countersuperflow instability into QT and its decay in two-component BECs at zero temperature by numerically solving the coupled Gross--Pitaevskii (GP) equations.
In contrast to QT in ${}^4$He thermal counter flow,
 the instability develops into QT without remnant vortices in our system.
 In the countersuperflow instability, quantized vortices are {\it intrinsically} nucleated in the bulk
 and then stretched by the momentum exchange between the two condensates [Fig. \ref{fig:dynamics} (a-d)].
 Eventually, quantized vortices of the two condensates get tangled with each other forming a binary QT [Fig. \ref{fig:dynamics}(e)].
 By calculating the time evolution of enstrophy of the binary QT,
 we investigate its isotropy and the similarity with classical turbulence. 
 We also demonstrate that the binary QT arising from the countersuperflow instability can be realized by the current experimental technique in cold atoms.

 Two-component BECs at low temperatures are described in mean-field theory \cite{Pethick} by the macroscopic wave functions $\Psi_j(t,{\bm r})=\sqrt{n_j(t,{\bm r})}e^{i\theta_j(t,{\bm r})}~(j=1,2)$
 with the particle densities $n_j$ and phases $\theta_j$,
 where the index $j$ refers to the $j$th component.
 The wave functions are governed by the coupled GP equations
 \begin{eqnarray}
i \hbar \frac{\partial}{\partial t} \Psi_j = \left(-\frac{\hbar^2}{2m_j}{\bm \nabla}^2+V_j+\sum_{k} g_{jk}|\Psi_k|^2\right)\Psi_j.
\label{eq:GP}
\end{eqnarray}
 Here $m_j$ and $V_j({\bm r})$ are the mass and the external potential of the $j$th component.
 The coefficient $g_{jk}=2\pi\hbar^2a_{jk}/m_{jk}$ represents the atomic interaction with $m_{jk}^{-1}=m_{j}^{-1}+m_{k}^{-1}$ and the s-wave scattering length $a_{jk}$ between the $j$th and $k$th components.
 Our analysis satisfies the conditions $g_{11}g_{22}>g_{12}^2$ and $g_{jj}>0$ that two miscible condensates are stable \cite{Pethick}.
  Note that, since there is no exchange of particles between the two components,
 the equation of continuity of mass is satisfied for each component: $\partial_t \rho_j+{\bm \nabla} \cdot {\bm J}_j=0$, where we use the mass density $\rho_j=m_jn_j$ and the mass flux ${\bm J}_j=\rho_j{\bm v}_j=\hbar n_j{\bm \nabla}\theta_j$ of the $j$th component.

 {\it Counter-superflow instability.} Before discussing trapped systems,
 we first elucidate the countersuperflow instability in an isolated uniform system with $V_j=0$.
 Because of Galilean invariance,  
 we may neglect the translational motion of the whole system without loss of generality:
 ${\bm P}_1+{\bm P}_2=0$ with ${\bm P}_j=\int {\bm J}_j d{\bm r}$.
 The wave function of the $j$th component in a stationary state may then be written as $\Psi^0_j=\sqrt{n_j^0}e^{\frac{i}{\hbar}(m_j{\bm U}_j\cdot {\bm r}-\mu_jt)}$
 with the chemical potential $\mu_j$ and the superfluid velocity ${\bm U}_1=-\frac{\rho_2}{\rho_1}{\bm U}_2$.

 The countersuperflow state is dynamically unstable when the magnitude of the relative velocity $U_{12}=|{\bm U}_1-{\bm U}_2|$ exceeds a critical value $U_c$.
 The critical relative velocity of the countersuperflow instability is essentially different from the critical velocity of the Landau instability \cite{Khalatnikov1},
 at which the frictionless flow becomes thermodynamically unstable.
 In the Landau instability, a superfluid exchanges momentum with a ``rigid'' environment (e.g., container wall or normal component),
 and the superfluid system is treated in the framework of a canonical ensemble.
 In contrast, the countersuperflow instability is purely an `'internal'' instability of the isolated system of two superfluids without any influence from the external environment.
 Thus the countersuperflow instability should lead to an exchange of momentum between the two components to reduce their relative motion, keeping the total momentum conserved in the uniform system.

 The momentum exchange must arise from a gradual excitation of internal motions, that is, from the appearance of excitations in the superfluids.
 The momentum exchange due to the excitations is investigated by solving the Bogoliubov--de Gennes (BdG) equations,
 which are obtained by linearizing the GP Eqs. (\ref{eq:GP}) with respect to a collective excitation $\delta \Psi_j=e^{\frac{i}{\hbar}(m_j{\bm U}_j\cdot {\bm r}-\mu_jt)} \bigl\{ u_{j}e^{i({\bm q}\cdot{\bm r}-\omega t)}-v_{j}^*e^{-i({\bm q}\cdot{\bm r}-\omega t)}\bigr\}$ around the stationary state.
The total momentum density $\delta {\bm J}$ carried by the excitation can be defined as $\delta {\bm J}=\delta {\bm J}_1+\delta {\bm J}_2$ with the change of the momentum density $\delta {\bm J}_j \equiv \hbar {\bm q}(|u_j|^2-|v_j|^2)$ of the $j$th component.
 The unstable modes, which trigger the countersuperflow instability, should satisfy the condition $\delta {\bm J}=0$ with $\delta {\bm J}_1=-\delta {\bm J}_2\neq 0$ due to the law of momentum conservation.
 Therefore, since the the countersuperflow instability is the dynamic instability triggered by the unstable modes with the imaginary part ${\rm Im}~\omega \neq 0$ \cite{Law},
 the amplification of the unstable modes exponentially enhance the momentum exchange.

{\it Nonlinear dynamics of the transition towards binary QT.} We demonstrate the characteristic nonlinear dynamics of the countersuperflow instability in an uniform system by numerically solving the GP Eqs. (\ref{eq:GP}) with $V_j=0$ in a periodic box.
 The parameters are set to be $m=m_1=m_2$, $g=g_{11}=g_{22}$, and $g_{12}=0.9g$, similar to the experimental situation discussed later.
 In the numerical simulation, the initial state is prepared by adding a small random noise to the wave function $\Psi^0_j(t=0)$ with $n_0=n^0_1=n^0_2$ and ${\bm U}_1=-{\bm U}_2$.
 Figure \ref{fig:dynamics} shows the time development of low-density isosurfaces with $n_1({\bm r})=0.05n_0$,
 where ${\bm U}_1=\frac{1}{2}U_{12}\hat{\bm x}$ with a unit vector $\hat{\bm x}$ along the $x$-axis.
 Because of the symmetric parameter setting, the $2$nd component behaves similarly to the $1$st component (not shown in Fig. \ref{fig:dynamics}).

 In the early stage of the dynamics,
 the amplification of the unstable modes creates disk-shaped low-density regions
 that orientate in the $x$ direction [Fig. \ref{fig:dynamics} (b)].
 The lowest density inside the disk region reaches zero, creating a local dark soliton \cite{Hamner10}.
 The soliton in the $j$th component transforms into a vortex ring \cite{B.P.Anderson} with a momentum  antiparallel to the initial velocity ${\bm U}_j$ [Fig. \ref{fig:dynamics}(c)].
 The distribution of the vortex rings can be determined from the dispersion relation $\omega({\bm q})$ obtained by diagonalizing the BdG equations;
\begin{eqnarray}
\hbar^2\omega^2=\epsilon^2+\frac{1}{4} \hbar^2 q_{||}^2U_{12}^2\pm\sqrt{\epsilon^2\hbar^2 q_{||}^2U_{12}^2+4g_{12}^2n_0^2\epsilon_0^2},
\label{eq:dispersion}
\end{eqnarray}
where we use $\epsilon^2=\epsilon_0(\epsilon_0+2gn_0)$, $\epsilon_0=\frac{\hbar^2}{2m}{\bm q}^2$, and ${\bm q}^2=q_{||}^2+q_{\bot}^2$ with $q_{||}=|{\bm q}\cdot \hat{\bm x}|$ and $q_{\bot} \geq 0$.
The instability condition $\omega^2<0$ is reduced to $\epsilon_-< \hbar q_{||}U_{12}/2 <\epsilon_+$ with $\epsilon_{\pm}=\sqrt{\epsilon_0[\epsilon_0+2(g\pm g_{12})n_0)]}$.
 This inequation determines the unstable region in the wave number space $(q_{||},q_{\bot})$ when $U_{12}$ exceeds the critical relative velocity $U_c=U_-$ with $U_{\pm}=2\sqrt{1\pm g_{12}/g}\sqrt{gn_0/m}$.
 The unstable modes have characteristic wave numbers $q_{||} \lesssim \frac{m}{\hbar}U_{12}$ and $q_{\bot} \lesssim \frac{m}{2\hbar}U_{12}$ for large relative velocity $U_{12}>U_+$.
 The length scale $\sim \frac{\kappa}{U_{12}}$ with $\kappa =2\pi \hbar/m$ then characterizes both the radii of the vortex rings and the intervals between the rings along ${\bm U}_{12}$ just after the vortex nucleations.
Thus, the vortex line density $l_v$ after the instability is roughly estimated to be $l_v \sim U_{12}^2/\kappa^2$,
 which is controllable by changing the relative velocity $U_{12}$.

\begin{figure}
\centering
 \includegraphics[width=.95 \linewidth]{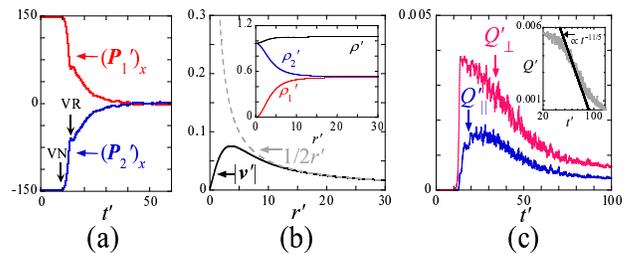}
 \caption{
 (a) Time evolution of momentum $({\bm P}_j)_x=\sqrt{gmn_0}({\bm P}_j')_x$ along the $x$-axis of the $j$th component.
 The two black arrows show the time around which the vortex nucleations (VN) and reconnections (VR) start.
 (b) Numerical plots of velocity $|{\bm v}|=\sqrt{gn_0/m}|{\bm v}'|$ and densities $\rho=mn_0\rho',~\rho_1=mn_0\rho_1',~\rho_2=mn_0\rho_2'$ (inset) as a functions of the distance $r=\hbar r'/\sqrt{gmn_0}$ from the core of the quasi-Rankine vortex in a stationary state.
 (c) Decay of anisotropic parameters $Q_{||}=q_0Q_{||}'$, $Q_{\bot}=q_0Q_{\bot}'$ and enstrophy $Q=q_0Q'$ with $q_0=(gn_0/\hbar)^2$ (inset).
}
\label{fig:enstrophy}
\end{figure}

The momentum exchange accelerates after the vortex ring nucleation [Fig. \ref{fig:enstrophy}(a)].
 The motion of the vortices then dominates the exchange.
 Since the momentum carried by a vortex ring increases with its radius,
 the radii of the nucleated vortex rings increase with time for the momentum exchange.
 This dynamics is similar to that of quantized vortices under the thermal counterflow of helium \cite{DonnellyBook},
 where the vortices are dragged by the mutual friction between the superfluid and normal fluid components.
 Our numerical simulations show that the ``mutual friction'' between the two condensates drags the vortices of each condensate under the countersuperflow.
 The ``mutual friction'' may be understood as an interaction between a vortex line and excitations.

 When the vortex rings become large,
the interaction between the vortex rings deforms the rings and vortex reconnections occur,
 which depresses the momentum exchange [Fig.\ref{fig:dynamics} (d) and Fig.\ref{fig:enstrophy}(a)].
 These effects make the vortex dynamics much complicated,
 leading to binary QT where vortices of both components are tangled with each other [Fig. \ref{fig:dynamics}(e)].
 Then the momentum exchange almost stops and each component has about zero momentum on average.

 {\it Statistical properties of the binary QT.}
 The interaction between vortices of different components,
 which is repulsive for $g_{12}>0$ \cite{Kasamatsu05}, becomes important when the mean intervortex distance is small.
 In fully developed binary QT, where the vortices of the two components are strongly correlated,
 it is plausible to regard the two condensates as a single fluid.
 We then introduce the total mass density $\rho=m_1n_1+m_2n_2$ and the total superfluid mass-current velocity ${\bm v}=({\bm J}_1+{\bm J}_2)/\rho$.
 Since the low-density regions along vortex cores of one component are filled with the other components due to the repulsive interaction between the components,
  the total mass density $\rho$ is leveled. 
Consequently, the velocity field ${\bm v}$ around the coreless vortex mimics that of the Rankine vortex \cite{Kund},
 where the velocity field  ${\bm v}$ is not irrotational and thus the vorticity field ${\bm \omega} \equiv {\bm \nabla}\times {\bm v}$ is continuous [Fig. \ref{fig:enstrophy}(b)].

It is convenient to investigate the statistical properties of the binary QT using the quantity $Q=\frac{1}{2V}\int {\bm \omega}^2d{\bm r}$ with the system volume $V$, which is called the enstrophy in classical turbulence \cite{Kund}.
 Analogizing with the discussion of helium QT in thermal counterflow \cite{DonnellyBook},
we define the anisotropic parameters 
\begin{eqnarray}
&&Q_{\parallel}\equiv \frac{1}{2V}\int {\bm \omega}_x^2d{\bm r},
\label{eq:Q_palallel}
 \\
&&Q_{\bot}\equiv\frac{1}{2V}\int ({\bm \omega}_y^2+{\bm \omega}_z^2)d{\bm r},
\label{eq:Q_vertical}
\end{eqnarray}
where $Q=Q_{\parallel}+Q_{\bot}$.
 If the tangle is isotropic, we have $Q_{\parallel}=Q_{\bot}/2$.
 Conversely if the tangle consists of curves lying only in planes perpendicular to ${\bm U}_{12}$, then $Q_{\parallel}=0$ and $Q_{\bot}=Q$.
 The time evolution of $Q_{\parallel}$ and $Q_{\bot}$ in the countersuperflow QT are plotted in Fig. \ref{fig:enstrophy}(c).
 The value $Q_{\bot}$ first appears with the vortex ring nucleation and increases due to the vortex line extension by the mutual friction keeping $Q_{\parallel}=0$.
 The vortex line {\it mixing} due to the vortex interactions and reconnections increases $Q_{\parallel}$ to $\sim Q_{\bot}/2$ when the turbulence becomes isotropic.
 Then the turbulence starts to decay with the enstrophy $Q$ keeping its isotropy $Q_{\parallel}\sim Q_{\bot}/2$.

We can directly analogize the binary QT with classical turbulence by investigating the time evolution of the decaying enstrophy.
 In a model for the decay of classical turbulence in a finite system \cite{Stalp},
 enstrophy decays like $Q(t)\propto t^{-11/5}$ in the early stage of the decay process.
 The enstrophy $Q$ of the binary QT follows this decay in the early stage [inset of Fig. \ref{fig:enstrophy}(c)],
 which means that the binary QT consisting of the quasi-Rankine vortices can mimic the classical turbulence as a turbulence of a single fluid.
 Such a viewpoint has not formerly been proposed and presents another direction for the analogy between classical turbulence and QT.
 To confirm this viewpoint, we have to find further similarities in different statistical quantities i.e. energy spectrum \cite{Kobayashi07}, velocity distribution \cite{White}, etc.
 A detailed analysis of binary QT will be reported elsewhere.

 {\it Counter-superflow QT in a trapped system.} Counter-superflow QT can be realized experimentally by employing a technique similar to that of Ref. \cite{Hamner10},
 where relative motion of the two condensates is induced by the Zeeman shift.
 We consider two-component BECs of ${}^{87}$Rb in the hyperfine states $|F,m_F \rangle=|1, 1 \rangle$ and $|2, -1 \rangle$,
 which can satisfy the miscible condition $g_{11}g_{22}>g_{12}^2$ by using the intercomponent Feshbach resonance \cite{Tojo}.
 We use an isotropic harmonic trap $V_j(t=0,{\bm r})=\frac{m}{2}\omega_r^2{\bm r}^2$ with $\omega_r=2 \pi \times 8$Hz
 and set the particle number to $N = 10^6$ for both components and the scattering lengths to $a=a_{11}=a_{22}=5.2~$nm and $a_{12}=0.9 a$.
 Since the full three-dimensional numerical simulation takes too much time and computer memory,
 we calculated a two-dimensional simulation considering the $z=0$ cross section of the condensates
 to investigate qualitatively whether the QT transition is observable experimentally.

\begin{figure}
\centering
 \includegraphics[width=.65 \linewidth]{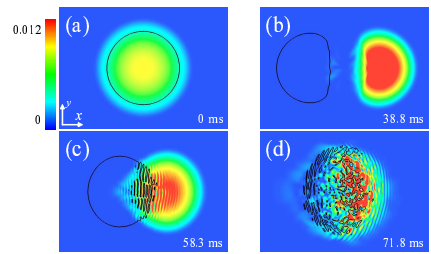}
 \caption{(Color online)
 Time evolution of the density profile $n'_1$ of the 1st component in the transition to binary QT from countersuperflow instability in a trapped system.
 The black lines represent the density contour $n'_2=0.004$ of the $2$nd component.
 The density $n_j$ is normalized as $n_j=n'_jR_T^4/(16a_h^6a)$ with $R_T=(15Na/a_h)^{1/5}$.
 The field of view is $29.2a_h \times 20.8a_h$ with $a_h=\sqrt{\hbar/(m\omega_r)}$.
 }
\label{fig:trap}
\end{figure}

 The condensates are first located at the minimum of the trap [Fig. \ref{fig:trap}(a)].
 A magnetic field gradient $-151$ mG/cm is applied at $t=0~$ms,
 which accelerates the two condensates in opposite directions along the $x$-axis due to Zeeman shifts.
 Although countersuperflow instability can occur in the region where the two condensates overlap,
 this collision process is not sufficient for the instability to grow into QT [Fig. \ref{fig:trap}(b)].
 However, this process is important for introducing sufficient seeds that develop into QT through the next collision process, described below.
 The field gradient is turned off at $t=11.7~$ms
 and the two condensates again return to the potential center ${\bm r}=0$ and collide there.
 Then, the countersuperflow instability forms a quasiperiodic density pattern consisting of many dark solitons [Fig. \ref{fig:trap}(c)].
 The solitons transform into vortex pairs,
 which corresponds to the vortex ring nucleations in three-dimensional space.
 The nonlinear vortex interaction leads to a turbulent state [Fig. \ref{fig:trap}(d)]
even if there is initially very small noise $\delta n_j({\bm r}) \sim 10^{-4}n_j({\bm r})$.
 Our simulation shows that the whole dynamic of the transition to binary QT from countersuperflow instability can be observed experimentally.
 A detailed analysis of trapped systems will be reported elsewhere.


\begin{acknowledgements}

 M.T. acknowledges the support of a Grant-in-Aid for Scientific Research from JSPS (Grant No. 21340104).
\end{acknowledgements}

\end{document}